# Generating single photons at GHz modulation-speed using electrically controlled quantum dot microlenses

A. Schlehahn,[1] R. Schmidt,[1] C. Hopfmann,[1] J.-H. Schulze,[1] A. Strittmatter,[1,a]
T. Heindel,[1,b] L. Gantz,[2] E. R. Schmidgall,[2] D. Gershoni,[2] and S. Reitzenstein[1]

[1]*Institut für Festkörperphysik, Technische Universität Berlin, 10623 Berlin, Germany*

[2]*The Physics Department and the Solid State Institute, Technion – Israel Institute of Technology, 32000 Haifa, Israel*

We report on the generation of single-photon pulse trains at a repetition rate of up to 1 GHz. We achieve this high speed by modulating the external voltage applied on an electrically contacted quantum dot microlens, which is optically excited by a continuous-wave laser. By modulating the photoluminescence of the quantum dot microlens using a square-wave voltage, single-photon emission is triggered with a response time as short as 270 ps being 6.5 times faster than the radiative lifetime of 1.75 ns. This large reduction in the characteristic emission time is enabled by a rapid capacitive gating of emission from the quantum dot placed in the intrinsic region of a p-i-n-junction biased below the onset of electroluminescence. Here, the rising edge of the applied voltage pulses triggers the emission of single photons from the optically excited quantum dot. The non-classical nature of the photon pulse train generated at GHz-speed is proven by intensity autocorrelation measurements. Our results combine optical excitation with fast electrical gating and thus show promise for the generation of indistinguishable single photons at high rates, exceeding the limitations set by the intrinsic radiative lifetime.

---
[a]  Present address: Abteilung für Halbleiterepitaxie, Otto-von-Guericke Universität, 39106 Magdeburg, Germany.
[b]  Author to whom correspondence should be addressed. Electronic mail: tobias.heindel@tu-berlin.de

With the prospect of realizing bright sources of non-classical light on a semiconductor platform [1,2], enormous research activities are under way worldwide [3,4]. In recent years, important steps have been taken, for instance, in the field of quantum communication where single-photon sources based on semiconductor quantum dots (QDs) can act as key building blocks for future technologies [5-7]. To further optimize efficiency, achievable single-photon rate and quantum optical quality of the generated photons in terms of high degrees of photon-indistinguishability [8], advanced nanophotonic devices with deterministically integrated QDs are very promising [9,10]. However, to achieve maximal repetition rates exceeding the GHz-range [11], a reduction of the lifetime of the QD states - typically ~ 1 ns in case of the InGaAs material system – is required [12,13]. Exploiting the Purcell effect in the weak coupling regime of micro- or nanocavity structures offers a solution [14], which, however, inherently suffers from a reduced spectral bandwidth. Alternative approaches to shorten the decay time of QD transitions are to manipulate the state occupation by electrical means. Such implementations accept the probabilistic nature of the emitting device (i.e some cycles of operation will not result in photon emission), but they are much more robust. For example, by applying a reverse bias charge carriers can be flushed out of the QD [15], or the quantum-confined Stark-effect can be exploited to tune the QD emission fast in- and outside the spectral detection window [16]. Such schemes typically require sophisticated pulse generators, delivering pulses with widths down to a few 10 ps, most probably hindering a cost-efficient and compact integration of such schemes in future.

In this work, we present an electrically contacted QD microlens, which, by applying a rather simple square-wave voltage, converts the emission of a single QD excited by a continuous-wave (CW) laser into a stream of triggered single photons with a characteristic emission decay time much faster than the electrical pulse-width itself, and also much shorter than the intrinsic radiative lifetime of the QD optical transition. We demonstrate, that the emission decay time can be reduced down to 270 ps, which corresponds to an improvement by a factor of 6.5 compared to the emitters' radiative lifetime (1.75 ns). This allows us to electrically trigger a single-photon pulse-train at 1 GHz repetition rate, even in the absence of cavity quantum electrodynamics effects.

The sample (see Fig. 1(a)) is a p-i-n diode grown via metal-organic chemical vapor deposition (MOCVD) with a low-density layer of InGaAs QDs located in the center of an intrinsic GaAs region. The microlenses are processed by means of 3D electron-beam lithography. After the lithography and etching of 4x4-arrays of lenses with variable size and shape (cf. Fig. 1(a), inset), a 100-nm-thick planarization layer of hydrogen silsesquioxane (HSQ) is spun onto the sample surface. The HSQ acts as isolation layer between the subsequently deposited semitransparent Ti/Pt contact layer (4/8 nm) and the semiconductor. Finally, to allow for electrical contacting, a large-scale planar Au pad is patterned via thermal evaporation. The whole sample is covered with an Au contact at the backside of the n-doped substrate. The current-voltage (I-V) characteristics of microlens-array as well as a representative micro-electroluminescence (µEL) spectrum of a microlens containing a single QD under 200 nA direct-current (DC) injection in forward direction is presented in Fig. 1(b). As expected from the pin-diode design, rectifying behavior with an onset voltage of about 3 V in forward direction is observed. The µEL spectrum shows distinct and narrow emission lines, which, due to the lack of any polarization dependence, are associated with charged QD transitions.

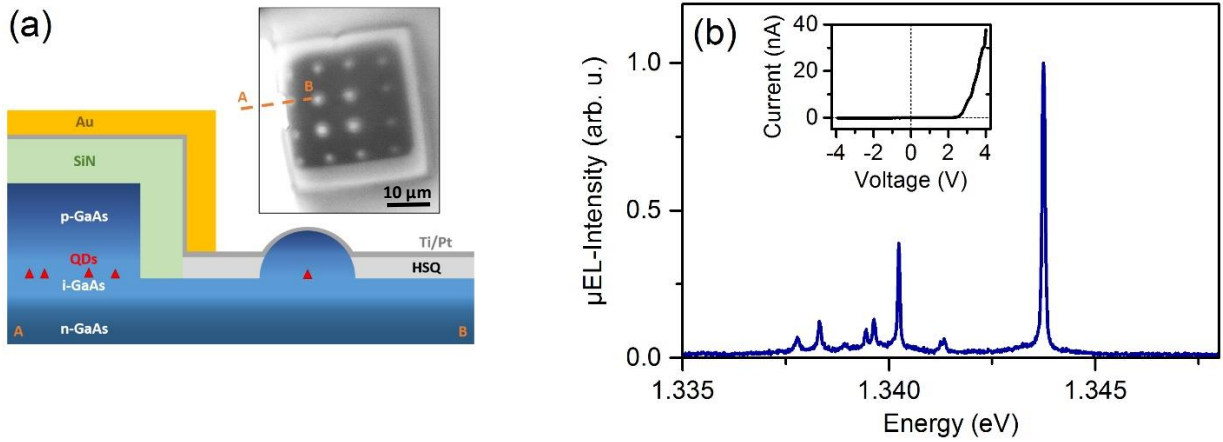

**Figure 1:** (a) Schematic cross-section of a p-i-n-doped QD-microlens sample. The microlens is contacted via a planar semitransparent Ti/Pt layer. Inset: Optical microscope image of a microlens array processed using 3D electron-beam lithography. The orange dashed line indicates the cross-section. (b) Electroluminescence spectrum of a single-QD microlens under 200 nA direct current injection (T = 20 K). Inset: I-V characteristics of the associated 4x4 microlens array.

For the generation of a single-photon pulse-train, the QD microlens is optically excited by a continuous wave (CW) diode laser emitting at 651 nm. The sample is mounted inside a He-flow cryostat (10 K) and can be electrically addressed by a micro-probe needle, allowing for convenient contacting the QD-microlenses. In order to electro-optically trigger the emission of single photons, a pulse generator featuring a maximum repetition rate of 3 GHz and variable pulse width, pulse amplitude and offset is used. Luminescence from the QD-microlenses is studied by a spectrometer, which comprises a double-grating monochromator and a charge-coupled device camera, enabling a spectral resolution of 25 µeV. For time-resolved measurements we use a Si-based avalanche photo diode (APD) with a timing resolution of 40 ps, which is directly coupled to the side-exit slit of the monochromator using a multi-mode optical fiber. Photon statistics are probed using a fiber-based Hanbury-Brown and Twiss (HBT) setup, comprising a 50:50 beamsplitter and two APDs with a timing resolution of about 300 ps.

Electro-optically triggered emission of a QD-microlens is presented in Fig. 2. To trigger the emission of single photons from the CW optically excited QD-microlens, electrical pulses with a width of 8.2 ns, a period of 12.2 ns - corresponding to a rate of 82 MHz - and an adjustable bias offset below the EL onset voltage (~2 V) are applied. Such an electrical pulse is indicated in the upper panel of Fig. 2(a). As a reference, constant PL emission under CW laser excitation at flat-band condition

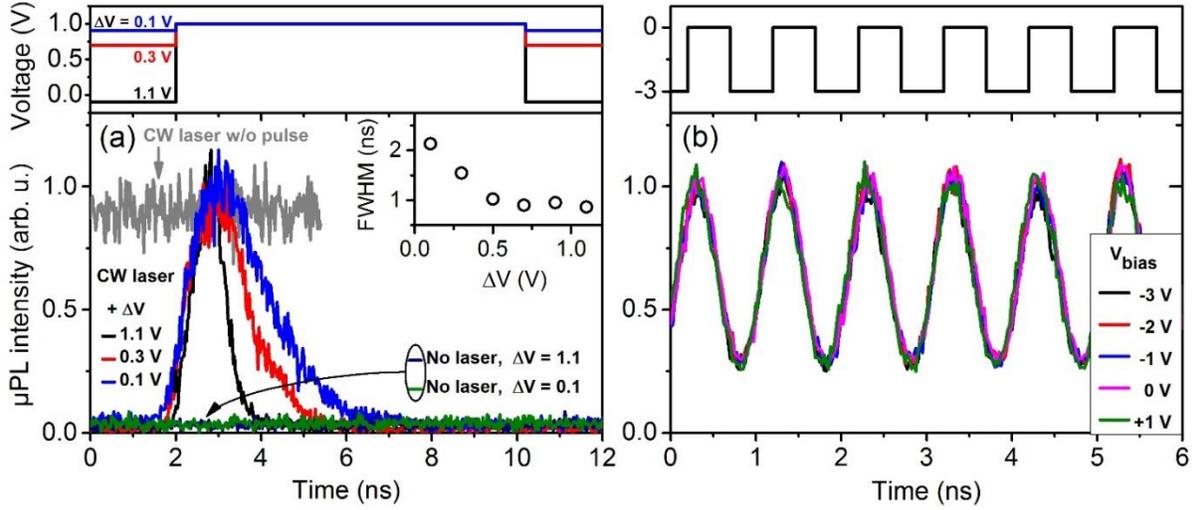

**Figure 2:** (a) Time-resolved photoluminescence from an electro-optically controlled QD-microlens, for various amplitudes of the electrical pulse. The applied electrical pulse (8.2 ns long, at a repetition rate of 82 MHz) with amplitude ΔV and bias offset $V_{bias}$ is schematically shown in the upper panel. The inset displays the resulting full width at half maximum (FWHM) of the emitted light pulse as a function of the pulse amplitude ΔV. (b) Upper panel: Schematic of an externally applied square-wave voltage (repetition rate of 1 GHz). Lower panel: Resulting pulsed emission for various bias voltages at constant pulse amplitude of 3 V.

($V_{bias}$ = 2 V) without applied electrical pulses is shown (cf. Fig. 2(a), grey curve). In the following, the pulse amplitude ΔV of the applied voltage is varied between 0 V (no pulses) and 1.1 V, by varying the bias voltage and fixing the maximum voltage level at +1 V (cf. Fig. 2(a), upper panel). Under such electro-optical excitation conditions, we observe a pronounced pulsed emission triggered by the rising edge of the electrical pulse, while emission is effectively quenched otherwise. Remarkably, even an amplitude ΔV as small as 0.1 V (blue line) is sufficient to trigger emission from the QD-microlens. Moreover, the optical pulse width $τ_{PL}$ depends on the electrical pulse amplitude, which allows for a reduction of the FWHM from 2.13 ns at 0.1 V amplitude (Fig. 2(a), blue line) to 0.86 ns at 1.1 V (Fig. 2(a), black line). In order to prove that the emission is indeed electro-optically triggered, rather than pulsed electroluminescence, two reference measurements without any optical excitation were performed for the smallest and largest amplitude (see Fig. 2(a)), resulting in no detectable signal. The fast periodic triggering of the emitter`s emission can be explained in terms of an abrupt depletion and refilling of the QD states via a fast tilt of the band structure. As the p-i-n junction is biased below the onset of EL, the optically generated charge-carriers are continuously expelled from the active area due to tunneling, except when recharging takes place at the rising edge of the applied pulse. Here, the fact that only the rising edge of the electrical pulse gates the emission of light, is attributed to the rectifying character of the diode. It is noteworthy, that in contrast to results obtained in Ref. [17], we observe no background emission in between the optical pulses for a sufficiently large amplitude, which can be an important advantage for e.g. applications requiring ultra-low background light-levels in pulsed experiments. The specific trigger mechanism observed in our QD-microlenses enables the QD to emit optical pulse trains in the GHz regime, which is demonstrated in the lower panel of Fig. 2(b). The applied 1 GHz square-wave voltage is schematically depicted in the upper panel of Fig. 2(b). The electrically triggered optical pulse train in Fig. 2(b) was recorded for various bias voltages between -3 V and +1 V at a constant amplitude of 3 V, revealing that in this case the quenching is not noticeably influenced by the offset within the limits of the pulse generator. A possible application for this operation scheme is, to realize a fast resonantly excited source of highly indistinguishable photons, which can be operated with standard CW tunable lasers [18]. In contrast, to achieve high degrees of photon-indistinguishability using standard current injection schemes is very challenging, due to the large intrinsic time jitter associated with above bandgap excitations.

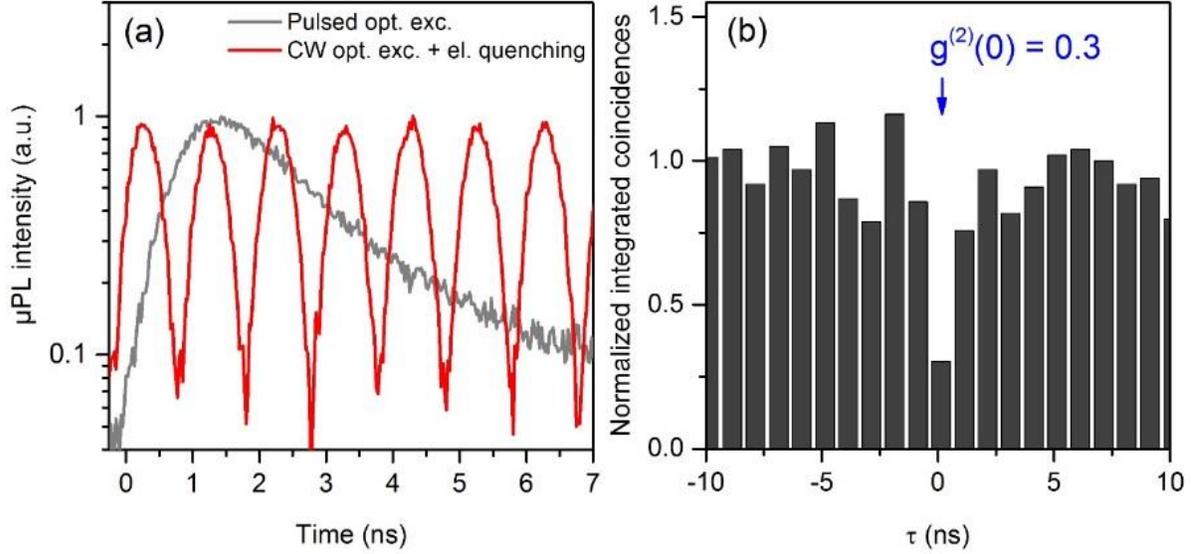

**Figure 3:** (a) Time-resolved photoluminescence emission resulting from a single QD exciton recombination under pulsed optical excitation without electrical pulsing (grey line) and under CW optical excitation with electrical modulation at 1 GHz repetition rate (red line). The photoluminescence signal duration is reduced by a factor of 6.5 due to the electrical depletion. (b) Temporal intensity-autocorrelation measurement of the photoluminescence described by the red line in (a). A $g^{(2)}(0)$ value of 0.3 is obtained.

As a proof-of-concept, an off-resonantly driven single-photon source operated at 1 GHz is demonstrated in Fig. 3. The repetition rate of triggered QD single-photon sources is usually limited by the radiative emission lifetime of the respective excited state, and can e.g. be enhanced by means of the Purcell effect in QD-microcavity structures [19]. To illustrate this aspect for the present system, time-resolved PL (TRPL) measurements were performed at a pulse repetition rate of 82 MHz without any electrical modulation at zero bias voltage. The corresponding PL transient is presented in Fig. 3(a) by the grey line. This trace is compared with the CW excited and electrically modulated emission at 1 GHz presented by the red line. From the TRPL data we extract a radiative emission lifetime of 1.75 ns, whereas the electro-optically modulated signal shows a decay time of 270 ps, which corresponds to a reduction by a factor of 6.5. Thus, in the present device design a significant enhancement of the maximal achievable repetition rate well above the radiative-lifetime-limited rate can be obtained without the need for complex microcavity structures, or sophisticated electronic pulse shaping. In order to prove the non-classical statistics of the light emitted within this fast optical pulse train, photon-autocorrelation measurements were performed using the fiber-coupled HBT setup. The corresponding coincidence histogram, which was recorded under the same electro-optical excitation parameters as the time-resolved measurements presented in Fig. 3(a), is shown in Fig. 3(b). Setting the time-bin-width (1 ns) to the inverse modulation speed, reveals a $g^{(2)}(0)$-value of 0.3, clearly demonstrating single-photon emission at a repetition rate of 1 GHz. This represents a promising starting point towards more sophisticated single-photon devices. For instance, since we use a planar semi-transparent metal layer to electrically address the microlenses, our contacting scheme can be easily combined with 3D in-situ cathodoluminescence-lithography techniques reported in Ref. [9]. These techniques can be applied to fabricate deterministic electrically controlled single-QD microlenses. In combination with resonant excitation schemes [18], this method becomes a powerful quantum device concept for the realization of highly efficient, ultra-fast sources of indistinguishable photons, and might even be completely integrated on-chip [20]. On the other hand, the fast electrically controlled depletion of QD charge-carriers observed in Fig. 3(a) is also highly desirable for advanced spin-control experiments using dark exciton states [21], and can increase the state-preparation-fidelity compared to optical depletion schemes used so far [22].

In summary, we presented fast electro-optical triggering of single photons from an electrically contacted QD-microlens. Our contact scheme makes use of fast electric field switching effects in order to overcome limitations on the single-photon-pulse-train generation, set by the intrinsic exciton lifetime. This allowed us to transform an off-resonantly excited, CW photoluminescence signal from a QD-confined exciton into extremely short optical pulses triggered by the rising edge of the applied electrical pulse where the optical pulse width is controlled by the voltage amplitude. The edge-triggering scheme leads to a reduction in the decay times of the optical pulses down to 270 ps, a factor of 6.5 in comparison to the intrinsic lifetime (1.75 ns) of the QD. It also enables the generation of single photons with $g^{(2)}(0) = 0.3$ up to an repetition rate of 1 GHz. Thus, our electrically contacted QD-microlenses may operate as highly-efficient sources of indistinguishable photons, operating in the GHz regime.

**ACKNOWLEDGMENTS**

This work was financially supported by the German-Israeli Foundation for Scientific Research and Development (GIF, Grant No. 1148-77.14/2011), the German Federal Ministry of Education and Research (BMBF) through the VIP-project QSOURCE (Grant No. 03V0630), and the German Research Foundation (DFG) within the Collaborative Research Center CRC 787 "Semiconductor Nanophotonics: Materials, Models, Devices".